\documentclass[12pt]{spieman}  % 12pt font required by SPIE;
\usepackage{amsmath,amsfonts,amssymb}
\usepackage{graphicx}
\usepackage{setspace}
\usepackage{tocloft}

\usepackage{upgreek}
\usepackage{xcolor}
\usepackage{booktabs}
\usepackage{array, ragged2e}
\usepackage{enumitem}
\usepackage{multicol}

\newcommand{\classname}[1]{\texttt{#1}}

\newcolumntype{R}[1]{>{\RaggedRight\arraybackslash}p{#1}}

\title{Mini-EUSO data acquisition and control software}

\author[a,*]{Francesca Capel}
\author[b, c]{Alexander Belov}
\author[d]{Giorgio Cambi{\`e}}
\author[d, e]{Marco Casolino}
\author[f]{Claudio Fornaro}
\author[b]{Pavel Klimov}
\author[d]{Laura Marcelli}
\author[e]{Lech Wiktor Piotrowski}
\author[e]{Sara Turriziani}

\affil[a]{Department of Physics, KTH Royal Institute of Technology, and The Oskar Klein Centre, Stockholm, Sweden}
\affil[b]{M.V. Lomonosov Moscow State University, D.V. Skobeltsyn Institute of Nuclear Physics, Moscow, Russia}
\affil[c]{M.V. Lomonosov Moscow State University, Faculty of Physics, Moscow, Russia}
\affil[d]{Istituto Nazionale di Fisica Nucleare - Sezione di Roma Tor Vergata, Roma, Italy}
\affil[e]{RIKEN, Saitama, Japan}
\affil[f]{Uninettuno University, Dipartimento di Ingegneria, Roma, Italy}

\cftpagenumbersoff{figure}
\cftpagenumbersoff{table} 
\begin{document} 
\maketitle

\begin{abstract}
We present the data acquisition and control software for the operation of Mini-EUSO, a space-based fluorescence telescope for the observation of extensive air showers and atmospheric phenomena. This framework has been extensively tested alongside the development of Mini-EUSO and was finalized ahead of the successful launch of the instrument to the ISS on August 22\textsuperscript{nd} 2019. The data acquisition, housekeeping and subsystem control is achieved using custom-designed front-end electronics based on a Xilinx Zynq XC7Z030 chip interfaced with a PCIe/104 CPU module via the integrated Zynq processing system. The instrument control interface is handled using an object-oriented C++ design which can be run both autonomously or interactively as required. Whilst developed for Mini-EUSO, the modular design of both the software and hardware can easily be scaled up to larger instrument designs and adapted to different subsystem and communication requirements. As such, this framework will also be used in the upgrade of the EUSO-TA instrument and potentially for the next EUSO-SPB2 NASA Balloon flight. The software and firmware presented herein are open source and released with detailed and integrated documentation.  
\end{abstract}

% Include a list of up to six keywords after the abstract

% Include email contact information for corresponding author
{\noindent \footnotesize\textbf{*}F. Capel,  \linkable{capel@kth.se} }

\section{Introduction}
\label{sect:intro}  

The study of ultra-high-energy cosmic rays (UHECRs) is limited by the rarity of events at the highest energies and the challenges posed by their detection. Their origin remains an open question which seems to become increasingly difficult to answer, even as more observations are acquired \cite{Kotera:2011kva, Batista:2019uf}. As proposed by the JEM-EUSO (Joint Experiment Missions - Extreme Universe Space Observatory) collaboration, the increased aperture of a space-based observatory would allow the detection of a higher rate of events, facilitating studies of the UHECR spectrum, arrival directions and composition \cite{TheJEMEUSOcollaboration:2015kz}. Taking the fluorescence detection concept to space would complement the observations of existing ground-based detectors and also be an asset to future multi-messenger studies.

The development of a space-based UHECR observatory presents obvious design challenges. While the fluorescence detection technique has been developed and used successfully on ground, the idea must be adapted to meet constraints on the size, mass and power requirements to make it possible to launch the project into orbit with current technology. Sensitivity to single photons is also necessary, requiring a high efficiency of the optical system and, in our design, photomultipliers with a high voltage power supply. In order to address these difficulties, the design is being developed and tested through a series of smaller-scale projects, the progress of which is well under way. Since 2013, the EUSO-TA instrument \cite{Abdellaoui:2018bq} has been installed on-site at the Telescope Array Project in Utah, USA. The EUSO-Balloon instrument \cite{Scotti:2016vgy} was launched in 2014, completing a $\sim$~8~hour flight over Timmins, Canada. Following this, the Tracking Ultraviolet Setup (TUS) instrument \cite{Klimov:2017fca} was launched into orbit in 2016 from Vostochny Cosmodrome and successfully collected data until December 2017. Most recently, in 2017, the EUSO-SPB (Super Pressure Balloon) instrument was launched from Wanaka, New Zealand, for a $\sim$~12~day flight over the Pacific ocean \cite{Wiencke:2017go}. These collective efforts have verified the space-based detection concept for UHECR-like signals \cite{Abdellaoui:2018cw}, uncovered a possible first UHECR candidate as seen from space \cite{Garipov:2019ux}, improved our understanding of the ultra-violet (UV) background relevant for extensive air shower observations \cite{Abdellaoui:2019ji} and confirmed the ability to detect other interesting atmospheric phenomena \cite{Klimov:2017fca}. More importantly, these instruments have also permitted the extensive testing of the hardware and software necessary for successful future experiments. 

We continue this development with the Mini-EUSO instrument, a small fluorescence telescope that is approved by the Russian and Italian space agencies and was successfully launched to the ISS from Baikonur Cosmodrome in Kazakhstan on 22\textsuperscript{nd} August 2019 on-board the Soyuz MS-14. With a time resolution of 2.5~$\upmu$s, referred to as the Gate Time Unit (GTU), and a spatial resolution of 6~km on the Earth's surface, the primary goal of Mini-EUSO is to map the Earth's atmosphere in UV with unparalleled detail. These measurements are crucial in order to properly understand the sensitivity of future orbital UHECR detectors and to design effective and robust trigger algorithms. With a high energy threshold of $\sim$~$10^{21}$~eV, we do not expect to observe UHECRs with Mini-EUSO. However, with its large annual exposure of $\sim$~15000~km$^2$~sr we will be able to provide competitive upper limits for a null detection. In addition to this, Mini-EUSO will also take advantage of its multi-level trigger algorithm to make scientifically interesting observations of transient luminous events, meteors, space debris and bioluminescence, and to search for nuclearites \cite{Casolino:2017jm}.

Compared to previous EUSO experiments, Mini-EUSO makes use of a new front-end electronics design that is more compact and powerful. Here, we present the data acquisition and control software developed for this system. A brief description of the instrument is given in Section~\ref{sect:instrument}, followed by an explanation of the software requirements and design in Section~\ref{sect:sw}. The integration and testing is described in Section~\ref{sect:testing} and conclusions are given in Section~\ref{sect:conclusion}. As Mini-EUSO is essentially one unit of a larger fluorescence telescope, this design will form the basis for planned future missions such as EUSO-SPB2 \cite{Scotti:2019cu}, K-EUSO \cite{Casolino:2017fg} and POEMMA \cite{Olinto:2017uq}. More generally speaking, the ideas presented here are relevant to any system based on a field-programmable gate array (FPGA) -- on-board computer interface to handle data acquisition along with the management of a range of subsystems. The software is released as open source \cite{MEsoftware19} with comprehensive documentation \cite{me_cpu_docs} which we hope will be a useful resource to those working on similar projects.

\section{The Mini-EUSO instrument}
\label{sect:instrument}

Here, we briefly summarize relevant concepts regarding the instrument design and operation. A more detailed overview can be found in the instrument white paper \cite{Capel:2017ig}.

\begin{figure}
\centering
\includegraphics[width=0.7\columnwidth]{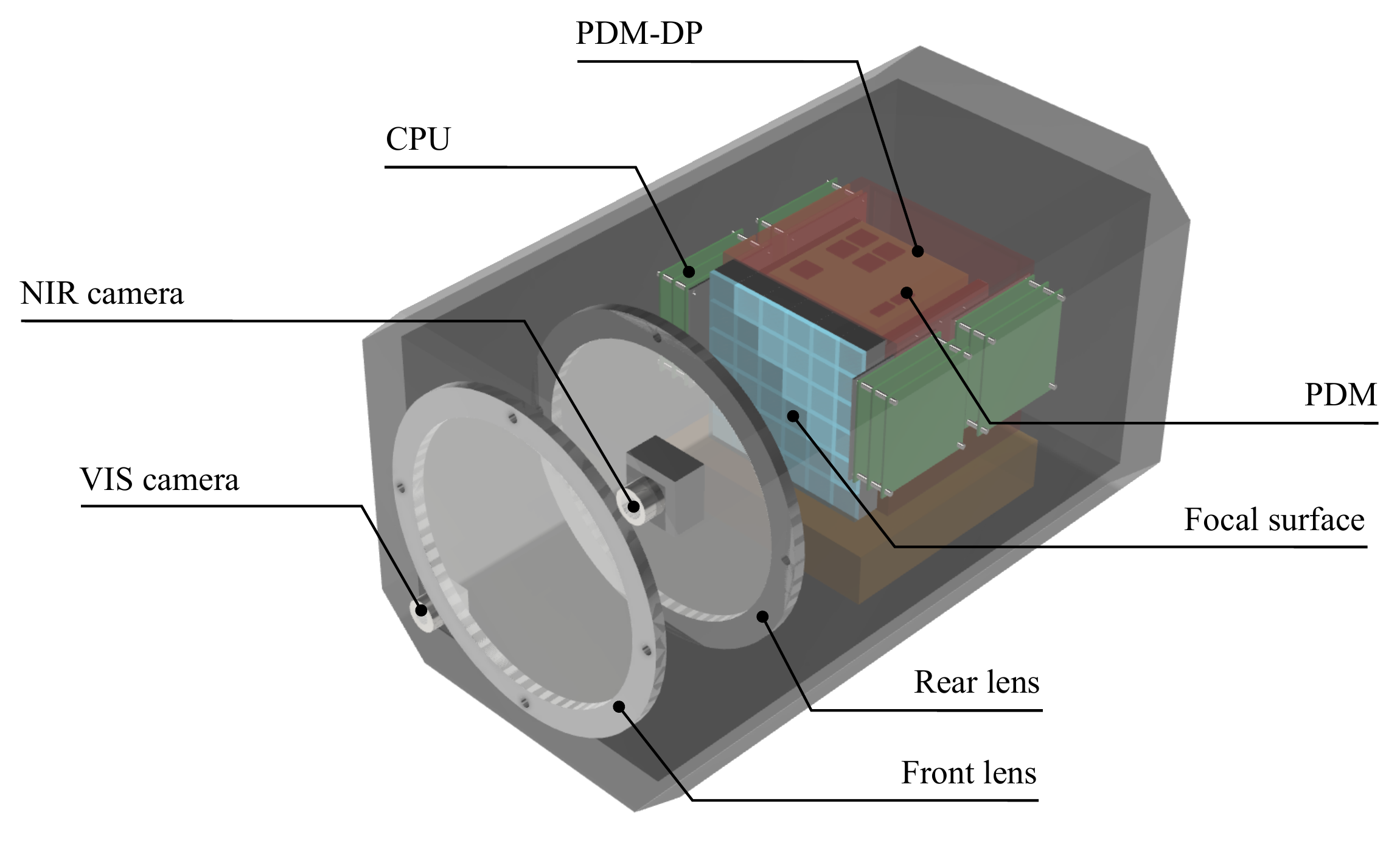}
\caption{Overview of Mini-EUSO showing the layout of the main subsystems. Incoming light is focused by the lenses onto 36 MAPMTs placed at the focal surface. Each PMT has 64 pixels, for a total of 2304 pixels, each with 6~km resolution on the Earth's surface. The data acquisition electronics are located in the back of the instrument behind the focal surface. The whole instrument is contained within $37~\times~37~\times~62$~cm$^3$.}
\label{fig:instrument}	
\end{figure}

\begin{figure}
\centering
\includegraphics[width=0.85\columnwidth]{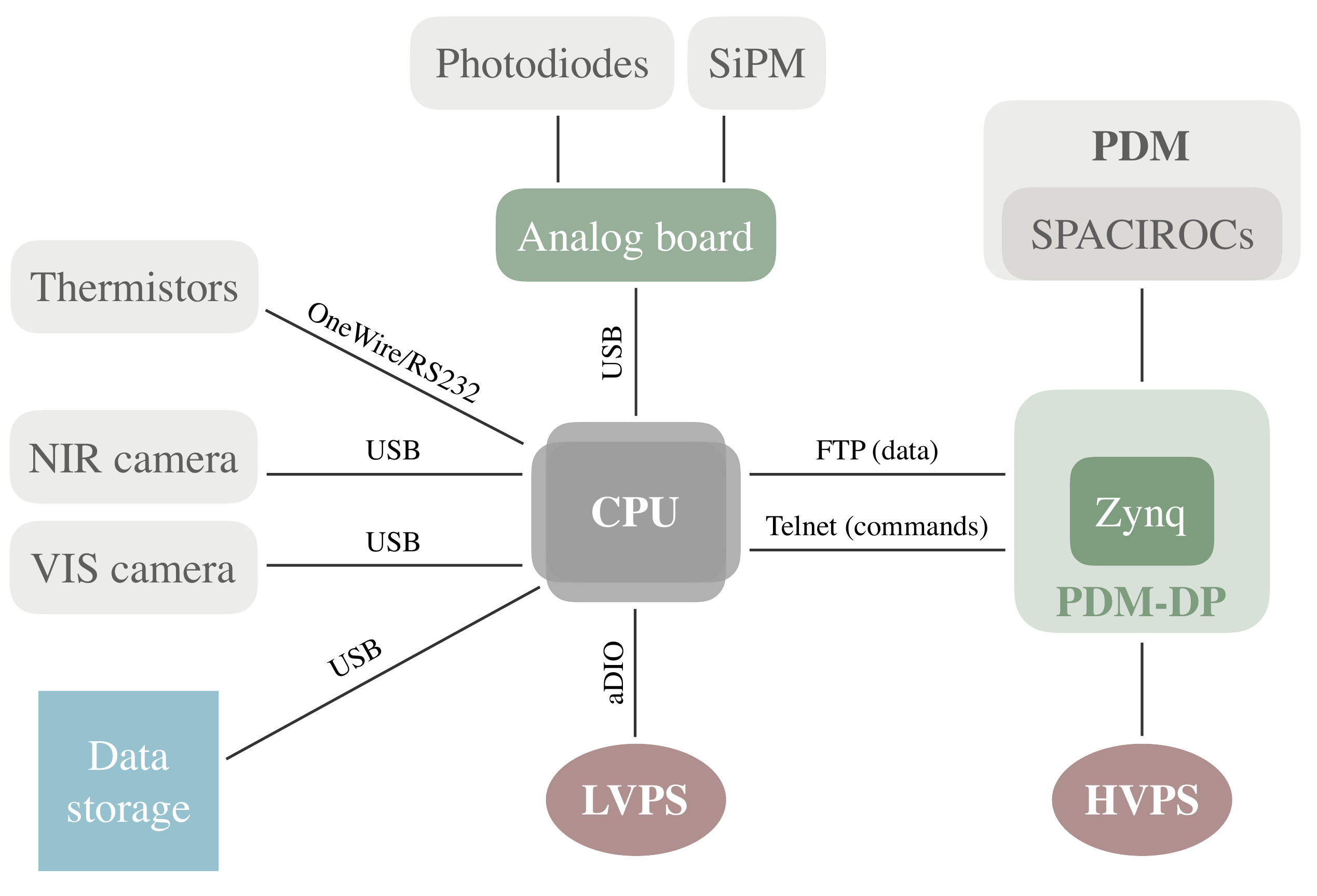}
\caption{The main interfaces of the CPU to the other subsystems are shown and labeled with their associated communication protocols. The CPU is powered by the low voltage power supply (LVPS) and can send signals over the built-in advanced digital input/output (aDIO) interface to switch all other subsystems on/off and monitor their status. Control of the HVPS and PDM is done by sending commands to the Zynq using a telnet connection over a TCP/TP link. Data is sent back over an FTP connection. The ancillary instruments have independent interfaces with the CPU. Sensors are shown in light grey, interface boards in green, power supplies in red and data storage in blue.}
\label{fig:interfaces}	 
\end{figure}

The Mini-EUSO instrument will be placed inside the ISS at a UV-transparent window on the Russian Zvezda module and will look down the nadir direction upon the Earth's atmosphere during local night-time. It consists of three main systems: the optics, the Photo Detector Module (PDM) and the data acquisition system, as shown in Figure~\ref{fig:instrument}. The optical system comprises two double-sided Fresnel lenses, a lightweight and compact solution to focusing light from a wide field of view (44$^\circ$) onto the PDM. The PDM is made up of the photomultiplier tubes and the front-end electronics. It has an array of 36 R11265-M64 multi-anode photomultiplier tubes (MAPMTs) supplied by Hamamatsu Photonics. These are covered with a BG3 UV filter so that the detector is sensitive to photons between 300-400 nm, the range in which the fluorescence lines of atmospheric Nitrogen peak. The MAPMTs are powered by a 1100~V Cockroft-Walton high voltage power supply (HVPS) \cite{Plebaniak:2017bh} and signals from the MAPMTs are processed and digitized with SPACIROC3 ASICs \cite{Blin:hi}. Signals from the SPACIROCs are passed to the data acquisition system which is made up of an FPGA board utilizing the Xilinx Zynq XC7Z030 chip \cite{Xilinx19} and a PCIe/104 CPU module (hereafter referred to as the CPU). The CPU is a commercial off-the-shelf CMX34BT module purchased from RTD Embedded Technologies, Inc \cite{rtd_cpu}. This is a single board computer based on a single-core Intel Atom E3815 1.46~GHz processor with 4~GB of DDR3 SDRAM, a 32~GB SATA flash drive and 6 USB 2.0 ports. The main interfaces of the CPU system are shown in Figure~\ref{fig:interfaces}.

The data acquisition system handles triggering, housekeeping, data storage and automated control of the whole instrument. Given its importance to the design of the software, we describe the hardware in some detail here, although further information can be found in Ref.\cite{Belov:2018ii}. As stated above, the data acquisition system is split into two main subsystems: the FPGA data processing (PDM-DP) and the on-board computer, or CPU. The PDM-DP interfaces to the SPACIROC ASICs with three Xilinx Artix7 FPGAs which perform data mapping and multiplexing. Data is then passed to the Zynq XC7Z030 system of programmable logic (PL, Xilinx Kintex7 FPGA), with an embedded processing system (PS, dual core ARM9 CPU). Most of the low-level data handling takes place in the Zynq system, including the implementation of the multi-level trigger algorithm \cite{Belov:2018ii}. This algorithm consists of an L1 trigger, which looks for UHECR-like signals on $\upmu$s timescales, a L2 trigger that looks for lightning-like signals on ms timescales and a L3 data type which is essentially a continuous readout providing information on transients with a timescale of 1~s, such as meteors. It should be noted that in this case, the trigger levels refer to different timescales, as opposed to more refined algorithms on the same timescale, as have been employed in previous EUSO instruments. This is because Mini-EUSO has a much larger field of view on the ground per pixel from its vantage point in Earth orbit compared to the past balloon experiments, and also has different science objectives concerning the observation of slower transients, such as lightning and meteors. The Zynq also handles the slow control of the ASICs and the high voltage power supply. The PDM-DP interfaces with the CPU using telnet and FTP connections. The CPU is responsible for the control and management of the data coming from the PDM-DP, as well as the control and data acquisition of all the other instrument subsystems, as highlighted in Figure~\ref{fig:interfaces}. Data is organised into packets and files to facilitate further processing offline. 

In order to keep the instrument running smoothly and to make complementary measurements, Mini-EUSO has a number of ancillary detectors and sensors installed which must be included in the software design. Near infrared (NIR, 1500-1600~nm) and visible band (VIS, 400-780~nm) cameras are mounted on the front of the instrument, outside of the optical system (see Figure~\ref{fig:instrument}), and are used to make complementary observations of the atmosphere during data taking \cite{Cameras19}. In order to determine when it is dark enough for the PDM to be switched on at high voltage, a set of photodiodes are installed around the edge of the focal surface. In addition to these, a 64-pixel silicon photomultiplier array (SiPM, C14047 series) from Hamamatsu Photonics will also be mounted at the edge of the focal surface in order to test this technology as a possible replacement for the MAPMTs in future instruments \cite{Renschler:vs}. The analog signals from the photodiodes and the SiPM are read out through an ATMegaS128-based analog board which is interfaced with the CPU. For housekeeping purposes, thermistors are used throughout the instrument to measure temperature and the status of power to all the subsystems is monitored through the low voltage power supply. 

As Mini-EUSO is designed to operate only during the local night-time, the maximum duty cycle is 50\%. We expect the actual duty cycle to be closer to $\sim$~10\% due to constraints from other activities on the ISS and the schedule of the astronauts. Assuming the 50\% duty cycle, the maximum data rate is then around $\sim$~3.2~TB/month, with the bulk of the data due to the main PDM acquisition, and the NIR and VIS cameras. This data is stored on USB flash drives with a capacity of 512~GB that have been sent up to the ISS along with the instrument. These drives are switched out on a roughly weekly basis by the astronauts as Mini-EUSO collects data. In this way, USB drives were chosen above SD cards for ease of handling and storage. 50 USB drives (25.6~TB) were sent up to the ISS along with Mini-EUSO, and these will be sent down to Earth and replaced roughly every 6~months, following the usual resupply schedule of the ISS. This procedure builds on past experience with the Sileye-3/Alteino experiment on-board the ISS \cite{Casolino:2002wk}. While there is no direct connection between the instrument and the ISS for the transfer of data, small ``quick-look'' datasets of around 1~GB/month will be prepared by the software. These datasets are then selected by the astronauts to be manually sent to  a mission control center using S- and Ku- band radio communications (2-4 and 12-18 GHz respectively) relayed though the Tracking and Data Relay Satellites (TDRS) by the ISS \cite{tdrss}. This ``quick-look'' dataset allows for checks and control of the Mini-EUSO functionality, as through this communication system it is also possible for the astronauts to change the software configuration file by loading it onto a USB drive, thus changing the behaviour of the software, as described in Section~\ref{sect:sw}.

\section{Software}
\label{sect:sw}

In this Section, we summarize the main concepts and considerations of the software design, the source code and documentation of which is released online as open source \cite{MEsoftware19}, \cite{me_cpu_docs} under the GNU General Public License, Version 3. The development lifecycle of the software followed the Agile development model \cite{agile_methods}, a dynamic and adaptive approach that is well-suited to the software development of a scientific project by a small team. In this way, the development phases of requirements, design, implementation and verification were repeated iteratively over short timescales to allow for a fast roll-out of a simple but functional version of the software that could be used throughout the development and testing of the instrument hardware. The first, basic version of the software was available in late 2016, with new functionality and subsystems added over the following years. In Section~\ref{sect:req}, we describe the main requirements that were defined throughout this process, Section~\ref{sect:imp} details the implementation and design choices to fulfil these requirements, and Section~\ref{sect:reuse} explains how the software can be reused and expanded upon. Details of the software verification and testing are given later in Section~\ref{sect:testing}. 

\subsection{Requirements}
\label{sect:req}

The main requirements are listed roughly following the chronological order in which their development unfolded, as different requirements were the most important in different phases of the software development lifecycle.

\begin{enumerate}
\item \emph{Interactive operation:} Prior to the installation of Mini-EUSO on the ISS, the instrument was integrated, tested and calibrated in the lab. During this phase, it was necessary to operate the software interactively with a flexible interface and variable number of subsystems.
\item \emph{Management of subsystems:} The CPU software is responsible for the management of all instrument subsystems, and so it must interface with each one independently. The main interfaces are summarized in Figure~\ref{fig:interfaces}, along with the protocols used. Details of the hardware interfaces, such as the various connectors and their pinout, are described in more detail in the online documentation \cite{me_cpu_docs}. The failure of one subsystem should not affect the operation of others from a software perspective. Due to hardware dependencies, in some cases this will be unavoidable. For example, should the PDM-DP fail, it will not be possible to control the high voltage. However, all standalone subsystems that have a direct connection to the CPU in Figure~\ref{fig:interfaces} should be able to operate independently.
\item \emph{Operation in different modes:} The ISS is in low Earth orbit at an altitude of $\sim$~400~km with a period of 93~min, meaning that it passes from local day to local night around every 46.5~min. To avoid damaging the MAPMTs as they operate in single-photon counting mode, Mini-EUSO can only acquire data from the PDM during local night-time. The instrument thus needs to have two distinct operational modes and make efficient use of the frequent day/night cycle to perform data reduction operations during the local day. 
\item \emph{Autonomous operation:} Once the Mini-EUSO instrument is installed, its only connection to the ISS will be for power (50~W) and grounding. This means that it will not be possible to continuously communicate with the instrument and it must operate autonomously and robustly following its launch. The main astronaut interaction will be to correctly position the detector, ensure it is powered, switched on, and to replace the data storage USBs at regular intervals. 
\item \emph{Configuration:}  We expect to be able to telemeter a small amount of ``quick-look'' data through the standard ISS TDRS communication system manually with the help of the astronauts. This channel will enable us to check and change the configuration of the software during the flight, so some irregular communication will be possible. In this way, the software design must be robust and automated, but also configurable in order to cope with any unforeseen complications. 
\item \emph{Flexibility and adaptation to future instruments:} The software must be designed to be flexible and adaptable, both during the development lifecycle of the Mini-EUSO project, and with the possible uses in future experiments in mind.
\end{enumerate}

The requirements can also be summarized in terms of the RAMS (Reliability, Availability, Maintainability and Safety) framework. Reliability concerns the ability of the instrument to operate as desired according to requirements 1, 3 and 4, and the management of subsystems in such a way as to properly handle failures, in accordance with requirement 2. Maintainability is the ease with which the software can be changed, updated, and repaired, as described by requirements 5 and 6. The availability is the percentage of overall uptime of the software, and is determined by the combination of reliability and maintainability. As there is no direct interface between the ISS systems and the software, there are no external safety requirements are imposed by the ISS. Safety is a concern for the instrument as a whole, but here we will focus on the software.

\subsection{Design}
\label{sect:imp}

\begin{figure}
\includegraphics[width=\columnwidth]{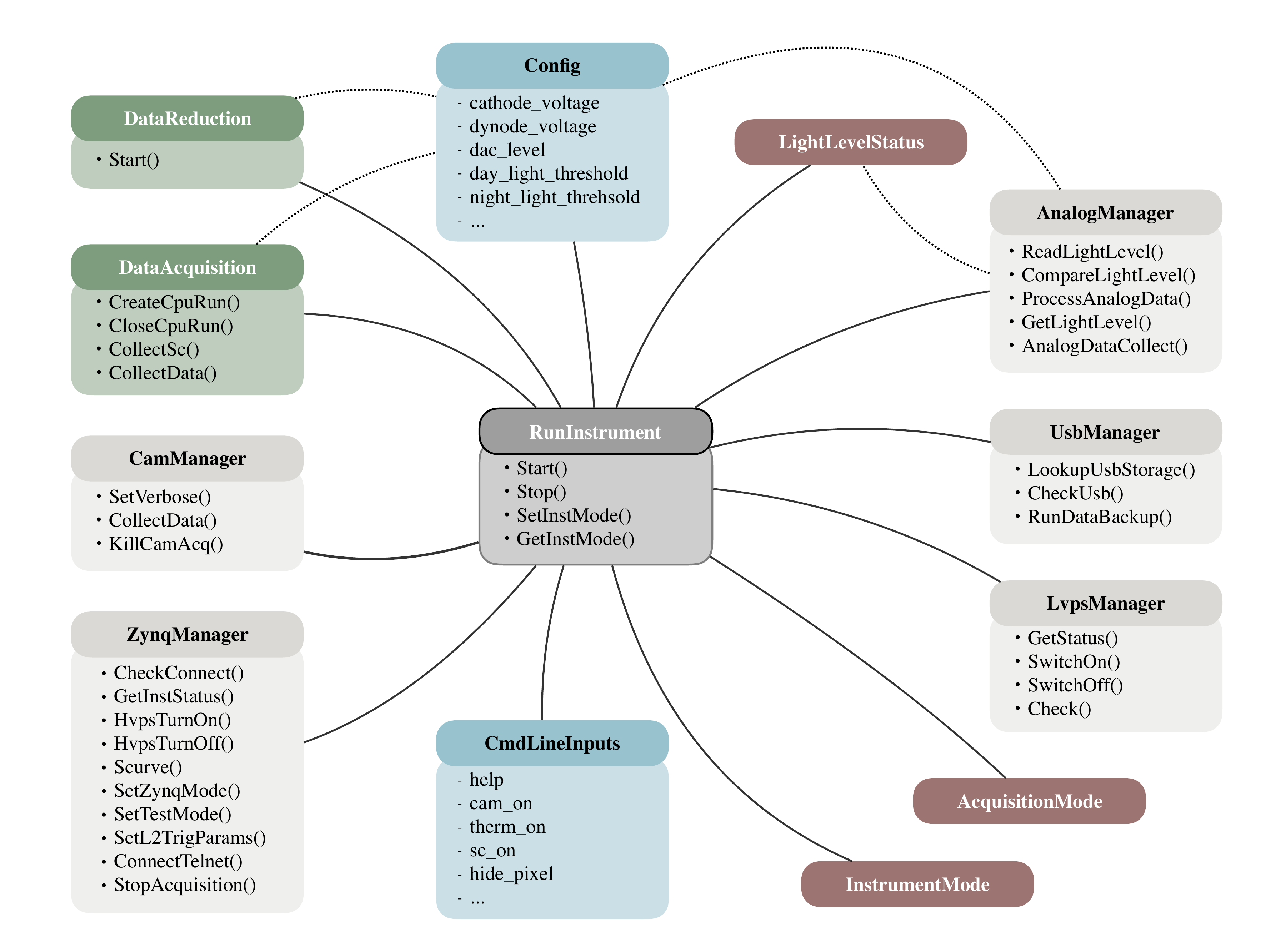}
\caption{Overview of the main class hierarchy in the Mini-EUSO software based on a first-order collaboration graph produced with Doxygen \cite{Doxygen}. The \classname{RunInstrument} class is responsible for the high-level interface to the instrument and makes use of the various subsystem manager classes (shown in light grey) and the \classname{DataAcquisition} and \classname{DataReduction} classes inheriting from the \classname{OperationMode} base class (shown in green). Configuration and command line inputs are stored in structures (shown in blue) which are then passed around the software as needed. Mutex guarded enumerations (shown in red) are used to store and share status information throughout the software. Key public class methods and struct members are shown to illustrate the functionality.}
\label{fig:sw_design}	
\end{figure}

In order to fulfil the requirements stated above, we decided to develop an object-oriented design in C/C++. The overall structure of the code is visualized in Figure \ref{fig:sw_design}. The \classname{RunInstrument} class handles the high-level control of the whole instrument such as checking the status of the subsystems, switching between operational modes and defining the autonomous operation sequence. The \classname{OperationMode} abstract base class is used to define the available operational modes. In the case of Mini-EUSO, these are \classname{DataAcquisition} and \classname{DataReduction}, depending on the night or day part of the ISS schedule respectively. The use of a base class here means that the signaling procedure for switching between modes is pre-defined, making it simple to add new operational modes into the flow of the main automated control sequence defined in \classname{RunInstrument}. Each subsystem is associated with its own class, resulting in a modular design which is easy to change and adapt for different instrument designs and requirements. For Mini-EUSO, the cameras, PDM-DP, data storage USBs, low voltage power supply, thermistors, photodiodes and SiPM all have separate subsystem classes. In addition to the main program structure, there are also classes for loading the configuration file (\classname{ConfigManager}), parsing the command line inputs generally used in interactive testing and calibration (\classname{InputParser}), logging with different levels to help debugging (\classname{logstream}) and other useful functions (\classname{CpuTools}).

\begin{figure}
\centering
\includegraphics[width=0.8\columnwidth]{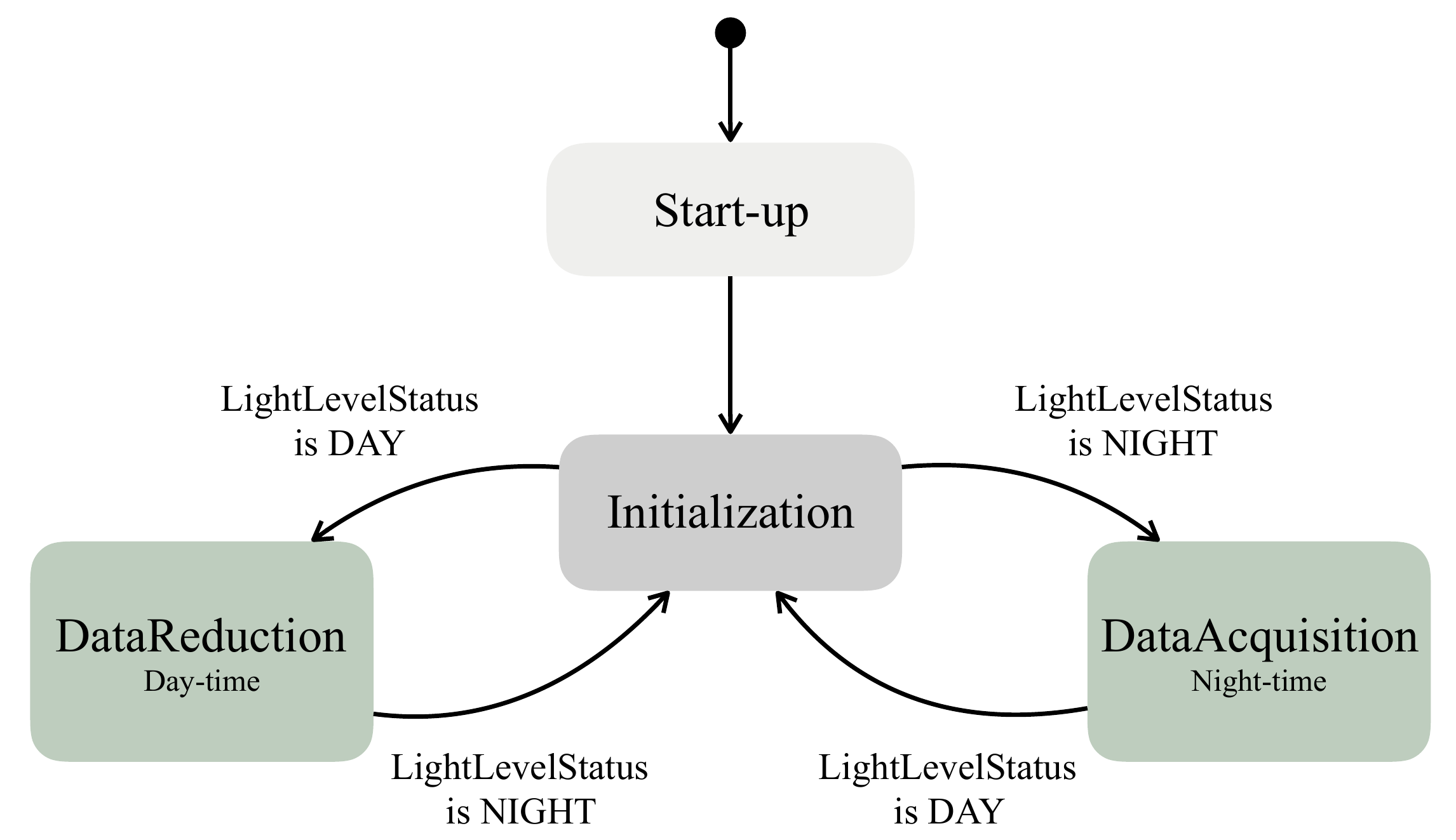}
\caption{The Mini-EUSO control loop. Upon switching on the instrument, start-up checks are performed on all the subsystems. Following this, the \classname{LightLevelStatus} is checked, determining the \classname{OperationMode}, which can be either \classname{DataReduction} or \classname{DataAcquisition}. The \classname{LightLevelStatus} is then polled regularly, and if it changes, a signal is sent to return to the initialization state and reset the \classname{InstrumentMode} accordingly.}
\label{fig:control_loop}	
\end{figure}

The high-level control flow of the program is implemented as a state machine such that once the instrument start-up procedures are completed, the instrument transitions to an \classname{InstrumentMode} state. These states are defined by an enumeration and realized with a corresponding class inheriting from \classname{OperationMode}. The control loop implemented for Mini-EUSO is summarized in Figure~\ref{fig:control_loop}. When the instrument is switched on, the boot and start-up procedures take around one minute to complete, after which the instrument enters an \classname{InstrumentMode} depending on the \classname{LightLevelStatus}, which is updated by reading out the photodiodes with the \classname{AnalogManager} class. Once an \classname{InstrumentMode} state is entered, the \classname{LightLevelStatus} is checked regularly. The period with which the \classname{LightLevelStatus} is polled is a configurable variable, but is set by default to 2~s. In this way, if the \classname{LightLevelStatus} changes, within 2~s the system will be notified, and the instrument will take around a few~s to exit its current state and transition back to the initialization state before moving to the required \classname{InstrumentMode}. If the instrument is being run autonomously and not interactively, this loop will then continue until the instrument is switched off. 

To fulfil the requirement that the subsystems should be able to operate as independently as possible in the case of a failure, the software performs a series of start-up checks when the instrument is switched on, handled by the \classname{RunInstrument::CheckSystems()} class method. These checks are used to determine the state of the instrument and inform its further operation. The checks and corresponding actions in the case of failure are detailed in terms of each subsystem in Table~\ref{tab:check_fail}. Due to the hierarchical dependencies of certain subsystems (e.g. the MAPMTs on the PDM-DP) it will not always be possible to ensure the success of the mission, but steps are taken to optimize the functionality where possible.

\begin{table}[h]
\begin{center}

\caption{Start-up checks for subsystems and corresponding procedures in the case of a failure. Certain systems are critical and in the case of a failure there is no action the software could take (E.g. if the LVPS fails, the CPU will not be powered in the first place). These subsystems are not considered here. }
\vspace{5pt}
\label{tab:check_fail}
\begin{tabular}{R{0.15\textwidth} R{0.4\textwidth} R{0.4\textwidth}}
\toprule
\textbf{Subsystem} & \textbf{Start-up checks} & \textbf{Failure procedure} \\
\midrule
NIR/VIS Cameras & 
\vspace{-20pt}
\begin{itemize}[leftmargin=*]
\itemsep-2pt
\item Check powered correctly using \classname{LvpsManager}
\item Try to launch cameras up to three times by catching and reacting to errors using the \classname{LaunchCam()} method of \classname{RunInstrument}
\end{itemize}
& Only switch on the working camera (if any) to save power. Continue nominal data collecting operations independently of collecting camera data. \\
\midrule
Thermistors &
\vspace{-20pt} 
\begin{itemize}[leftmargin=*]
\itemsep-2pt 
\item Check responding using the \classname{GetTemperature()} method of \classname{ThermManager}
\end{itemize}
& If not responding, write a default value to the output instead that will be recognized in the data. The temperature is not used elsewhere in the software, only as complementary information to help debug on the ground. \\
\midrule
Analog board & 
\vspace{-20pt} 
\begin{itemize}[leftmargin=*]
\itemsep-2pt
\item Check powered correctly using \classname{LvpsManager} 
\item Check responding using the \classname{AnalogDataCollect()} method of \classname{AnalogManager}
\item Check value of the \classname{LightLevelStatus} 
\end{itemize}
 & If not responding, put the instrument in \classname{DAY} mode to avoid damaging the MAPMTs and give a chance to debug on ground. If values are not as expected, the separate day and night thresholds can be configured by uploading a new configuration file.  \\
\midrule
Data storage &
\vspace{-20pt} 
\begin{itemize}[leftmargin=*]
\itemsep-2pt 
\item Check the number of USB storage devices connected using the \classname{LookupUsbStorage} method of \classname{UsbManager}
\end{itemize}
  & If no storage devices are connected, store data on the local hard drive, if it is not full. \\
\midrule
PDM-DP & 
\vspace{-20pt} 
\begin{itemize}[leftmargin=*]
\itemsep-2pt 
\item Check powered correctly using \classname{LvpsManager}
\item Check telnet and FTP connection using the \classname{CheckStatus} method of \classname{RunInstrument}, as well as the HVPS status
\end{itemize}
 & If the PDM-DP is not responding, it will not be possible to control or collect data from the PDM. In this case, housekeeping data and log messages will be stored in files to allow for debugging of the instrument from the ground. \\
\bottomrule 
\end{tabular}
\end{center}
\end{table}

Any instrument with a number of subsystems will need to handle multiple tasks simultaneously. We achieve this by launching multiple processes and threads which operate asynchronously and communicate with each other via signaling and common memory. Each thread is assigned a specific task, and has a mechanism for ending or joining cleanly upon request. Any class members which are accessed by multiple threads are protected by a mutex to avoid concurrent access. All multithreading is implemented in accordance with the C++11 standard. Under nominal operations, after performing system checks and determining the operational mode, \classname{RunInstrument} launches two background processes: instrument monitoring and data backup. Instrument monitoring checks the light level of the photodiodes periodically to determine when it is time to switch from \classname{DataAcquisition} to \classname{DataReduction} mode and vice versa. In order to be robust to fluctuations, a moving average is taken over all the photodiode measurements and separate thresholds are used for the day~$\rightarrow$~night and night~$\rightarrow$~day transitions. When the configured light threshold is crossed, all other spawned processes and threads are signaled to end and the main program alone enters the desired mode. The data backup thread simply looks for new data files on the USBs and copies them over to a separate USB continuously in the background using rsync. When an operational mode is entered by the main program, this will in turn launch processes and threads to handle the necessary tasks. For example, \classname{DataAcquisition} launches a process to acquire data from the cameras and separate threads to handle the acquisition from the housekeeping sensors and PDM-DP as well as to watch for and process the incoming data. The \classname{SynchronisedFile} and \classname{Access} classes are used to handle asynchronous file writing so that data packets can be efficiently gathered and stored from all subsystems simultaneously.

When the standard data acquisition is running, the maximum CPU load is $\sim$~40\%. Despite the fact that the operating system (OS) is standard linux and not a Real-Time (RT) OS, we do not need to be concerned with data loss due to temporary CPU overhead. The main PDM data acquisition is carried out through the Zynq system, which is essentially an RT component decoupled from the CPU. This decoupling means that a non-RT OS is suitable, allowing for more flexible development of the software. A similar design was implemented for the on-board software of the EUSO-SPB mission~ \cite{Fornaro:2018jg} and subsequently tested during the balloon flight.

The main data acquisition proceeds via interfacing the CPU with the PDM-DP. Commands are sent from the CPU to the Zynq PS using a telnet connection over a TCP/IP link and implemented in the code using socket programming. The Zynq PS then configures the Zynq PL to carry out the desired functions, from switching on the high voltage, to checking the instrument status and data gathering. The Zynq system has a variety of data acquisition modes including different trigger combinations and also diagnostic options for integration and testing purposes. Under nominal operations, data is acquired using the multi-level trigger algorithm implemented in the Zynq PL. We made use of high-level synthesis \cite{XilinxHLS} to design and implement a trigger logic which can be configured in real-time using commands sent from the CPU to the Zynq~\cite{HLSzynq19}. This means that both the software and FPGA firmware can be adapted to deal with unexpected observational conditions in-flight, or indeed to optimize the same trigger algorithm for different instruments. During data collection, the PDM-DP periodically sends files to the CPU over an FTP connection, with the Zynq PS as the server and the CPU as the client. These data files are then processed asynchronously by the CPU along with information from the other subsystems and stored on the USBs.

Whilst the software is designed to run autonomously once the instrument is powered, it can also be run in an interactive mode by passing command line inputs to the main program. In this way, simple execute-and-exit commands can be run, such as switching on the high-voltage or checking the instrument status, or even whole acquisition sequences, such as taking a fixed number of data packets from the Zynq in a certain mode, with all other subsystems gathering data. These commands are described in detail in the online documentation \cite{me_cpu_docs}, as well as standard test procedures to verify the basic functionality of the instrument. This flexibility has been essential in the development and testing of both the software and hardware of Mini-EUSO. The \classname{RunInstrument} class can also handle standard interrupt signals (e.g. SIGINT) properly in order to stop all operational threads cleanly and switch off the high voltage to the MAPMTs before exiting the main program. 

\subsubsection{Data format}

\begin{figure}
\centering
\includegraphics[width=0.8\columnwidth]{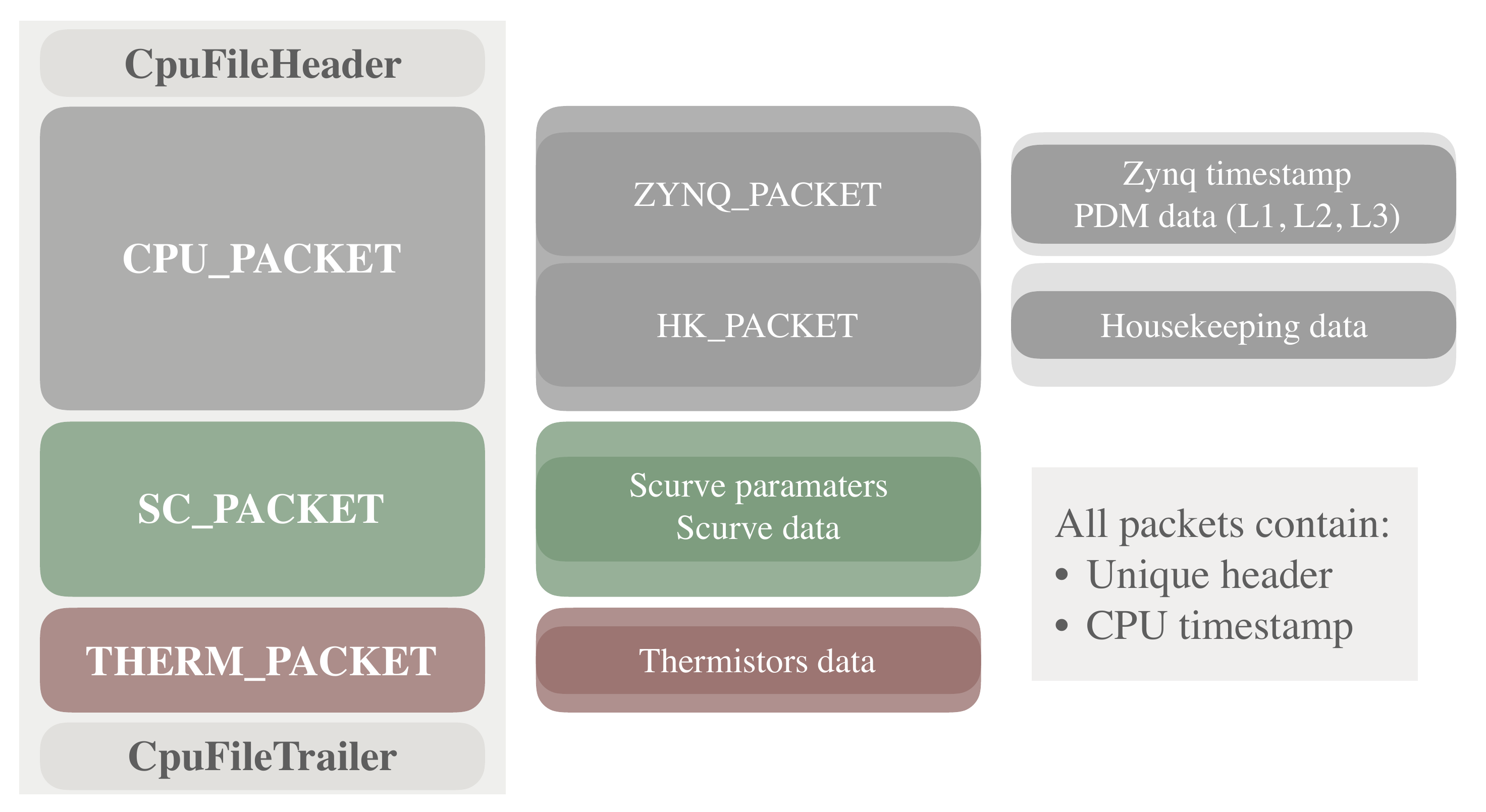}
\caption{The Mini-EUSO data format. A file is shown with a few different packets to demonstrate the hierarchical structure. Typical files will contain around 25 packets which are written asynchronously for efficiency. Data from the cameras will be stored separately.}
\label{fig:data_format}	
\end{figure}

The data format was designed to retain similarity with that of previous experiments to allow the use of existing tools within the EUSO collaboration for offline data analysis, but also to have the flexibility to cope with the inevitable changes that will come with future developments. Data are organised into separate packets which can be easily recognised by their headers. These packets are gathered together in files within a hierarchical ``matryoshka'' style structure, containing more and more detailed information, as shown in Figure~\ref{fig:data_format}. The main packet is the \classname{CPU\char`_PACKET}, which packages together information from the PDM-DP with relevant housekeeping and high voltage status data from the time of readout. Other packets include \classname{THERM\char`_PACKET} containing temperature information which is read out asynchronously with a low frequency for efficiency, the \classname{SC\char`_PACKET} for ``S-curves'' (photoelectron count rate change with digitization threshold level, used to characterize MAPMTs and ASICs) and \classname{HV\char`_PACKET}  which provides complementary information on the high voltage status during a data acquisition run. The packet structure means that it is easy to add and remove different information from the data format without requiring many low-level changes to the code. 

Each data file contains around 25 packets and is bounded by a separate header and trailer which contain further information on the commands used to start the acquisition run and a 32-bit cyclic redundancy check. The exact file size is variable due to the asynchronous nature of the readout, but a typical file size is $\sim$~118~MB for a standard acquisition. These binary data files will later be converted to the CERN ROOT \cite{ROOT} file format for offline data analysis using a library to identify packets via their unique header tags and package them into ROOT TTrees \cite{Piotrowski:2017gv}. As mentioned in Section~\ref{sect:instrument}, we expect a maximum data rate of $\sim$~3.2~TB/month. If the USB storage devices become full before being switched out by the astronauts, the software will simply stop writing data until the USB storage device is replaced. The astronaut training and planned Mini-EUSO activities are designed to avoid such an eventuality, with USB drive replacement taking place in accordance with the expected data rate.

\subsubsection{Data Reduction}

Once the light level measured by the photodiodes exceeds the threshold for the safe operation of the MAPMTs, the instrument switches to the \classname{DataReduction} operational mode; the high voltage power to the MAPMTs is ramped down and data acquisition threads are exited. During the local day, the CPU processing power is focused on the filtering and compression of data from the previous night-time run, in order to produce the quick-look sample to be downlinked to ground during the flight. We expect a rate of $\sim$~1~GB/month of data (after filtering and compression) to be transferred to ground. Initially, we plan to use this small data budget for the transfer of diagnostic and housekeeping information on the various subsystems at early stages in the flight of Mini-EUSO, in order to verify the functionality of the instrument and tweak its configuration. As the operation of the instrument progresses, we plan to send packets containing interesting candidate events for the L1 and L2 output data levels, filtered using dedicated algorithms. We expect the bulk of the data to be returned to Earth during the lifetime of the experiment on the ISS, in accordance with the ISS resupply schedule every few months. However, the data reduction will be used to filter out particularly interesting events which would then be prioritized for the smaller, more frequent data transfer using the ISS communications systems.

In addition, the ancillary cameras will continue to operate at a reduced sample rate during the day with a continuous readout for atmospheric monitoring, accompanied by housekeeping data read out at the same frequency of $\sim$~0.5~Hz. This reduced data sample will be acquired and stored in the same manner as the main data acquisition during the local night.

\subsection{Reusability}
\label{sect:reuse}

Other subsystems can be included into the software design by adding a new \classname{SubsystemManager} class. There is currently no abstract base class or template for subsystems as the nature of the different subsystems in Mini-EUSO is so different. However, should one want to add many similar subsystems, this would be the preferred implementation. The new \classname{SubsystemManager} class can then be instantiated and used by the \classname{RunInstrument} and \classname{OperationMode} classes, which provide a flexible interface for managing different subsystems and communicating between them. The main requirement for a \classname{SubsystemManager} class is that if it launches any threads, there should be a clearly defined class method for cleanly exiting the thread. This way, the \classname{RunInstrument} can properly handle switching between different \classname{OperationMode}s. New operational modes can also be defined by inheriting from the \classname{OperationMode} abstract base class, as demonstrated by the existing operational modes, \classname{DataAcquisition} and \classname{DataReduction}. The \classname{OperationMode} base class has built in \classname{Notify()} and \classname{Reset()} switches which can be used to exit and initialize the mode. It also carries instances of \classname{AnalogManager} and \classname{ThermManager} so that basic housekeeping tasks can be performed in any mode. New \classname{SubsystemManager}s could also be added, if desired, in a similar fashion. Once a new \classname{OperationMode} has been defined, its implementation, state and transition requirements must be specified in \classname{RunInstrument}.

The design of larger EUSO instruments in the future will involve multiple PDMs, resulting in a larger optics, focal surface and sensitivity to lower energy UHECRs. The idea is to have a distributed network of individual PDM-DPs, one for each PDM, connected by high-speed ethernet and controlled by a central CPU-like system. In order to scale up this software design to a multi-PDM instrument, we would have to introduce another level of hierarchy into the current design.  A \classname{PdmManager} class would be added to handle the lower level interface to each PDM and PDM-DP individually, with an \classname{Instrument} class storing the number and configuration of PDMs for a particular instrument. With these additions, the high-level control of \classname{RunInstrument} could be adapted to handle a large number of PDMs. Further work would also be needed to synchronize the data flow from the various PDMs, but this could be done by building on the current acquisition system using the \classname{DataAcquisition}, as this would largely be handled by the PDM-DP, so the job of the CPU software would be to package the data into files conveniently.

\section{Integration and testing}
\label{sect:testing}

\begin{figure}
\centering
\includegraphics[width=0.6\columnwidth]{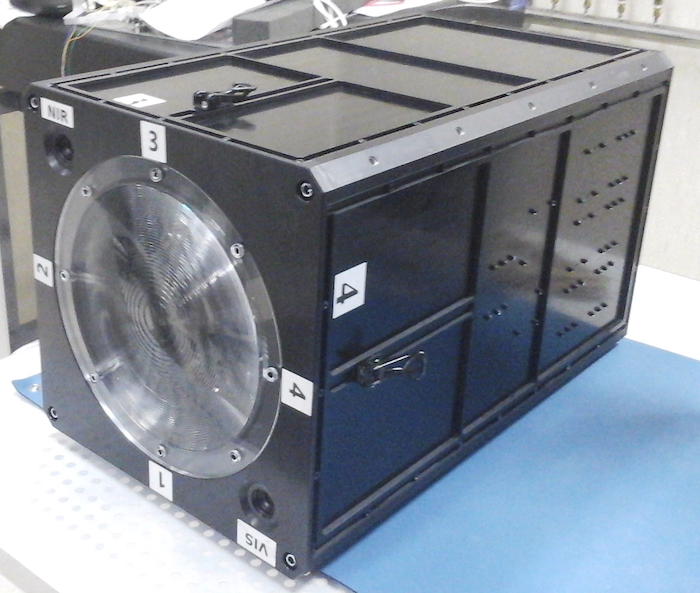}
\caption{The Mini-EUSO flight model in the clean room at the University of Rome Tor Vergata during recent spaceflight acceptance tests.}
\label{fig:me_integration}	
\end{figure}

The data acquisition and control software for Mini-EUSO has been developed and extensively tested throughout the integration of both the engineering and flight models and subsequent qualification tests at the RIKEN research institute in Tokyo, Japan, and the University of Rome Tor Vergata, Italy. We have verified the correct operation and robustness of the software both on simulated data passed directly to the front end electronics \cite{hardwaretestbench} and in the laboratory. End-to-end tests in the TurLab facility at the University of Turin, Italy \cite{Bertaina:2015hha, Bertaina:ur} in February 2018 confirmed the correct operation of the software and trigger algorithms in simulated orbital conditions. Additionally, Mini-EUSO has also been tested using observations of stars and planets in the night sky from the ground \cite{Cambie:wb, Bisconti:2019wq}. A picture of the integrated flight model is shown in Figure~\ref{fig:me_integration}. The varied functional testing process allowed us to test both the interactive and automated code, and tests in the lab were designed to give good coverage of the whole software under both nominal and overloaded cases. To evaluate the software availability, we ran long, unsupervised acquisitions with the instrument exposed to a controlled light level in a black box to simulate the conditions in orbit. This allowed us to remove or handle any of the observed catastrophic failures, resulting in an expected availability of $>$~95\% for the instrument as a whole. Prior to the launch of Mini-EUSO, it was required to pass specific spaceflight qualification tests regarding the safety and functionality of the instrument. Relevant for the software, the functionality test required the instrument to be run autonomously requiring only power, grounding and to be switched on. The expected data files should then be stored on the USB flash drive and checked to contain the expected information by converting them to ROOT files and checking using the EUSO data visualization software, ETOS \cite{Piotrowski:2017gv}. 

In addition to functional testing, software quality assurance measures were taken at all stages of the design and implementation of the software. During the design phase, ideas were discussed and reviewed by the Mini-EUSO team and the JEM-EUSO Collaboration as a whole to ensure the proper treatment of all the subsystems and optimization of the design for the science goals of the project. The implementation of the software was standardized by following the Google Style Guide for C/C++\cite{google_style_guide} and Codacy's static analysis tools\cite{codacy} integrated into the software's GitHub repository. Codacy assigns a grade from A to F, with A being the highest grade, based on the number of issues identified for every thousand lines of code. We maintained a Codacy grade of A in order to ensure high quality code. Travis CI\cite{travis_ci} was used for continuous integration, with automatic cloud-based builds set up for every new commit or pull request to the main repository. A successful build was required in order to change the code on the master branch. Travis CI can also be set up to run integrated unit testing, but unfortunately due to the custom hardware required to fully test the code it would be necessary to design a simulator in order to implement this automated, cloud-based testing apprach. This was not done due to the time constraints of the project and the Mini-EUSO launch, but unit tests of all subsystems were written into \classname{RunInstrument::DebugMode()} and run manually when testing in the lab. Automated testing as part of the continuous integration using Travis CI and GitHub is a possibility that will be explored in the future development of the software. 

The software should be compatible with all linux-based operating systems but has been primarily developed running on Debian 8 (amd64) and tested on both the CMX34GS and CMX34BT series PCEe/104 CPU modules. Preliminary testing of the software has also been successfully completed on Ubuntu versions 16.10 and 18.04 (amd64, LTS). In order to run on other, non-linux operating systems, changes would need to be made to the filesystem monitoring used in the data processing part of the code.  At the time of writing, version 8.1.1 is released and the software development now proceeds with development for EUSO-TA and future instruments, following the successful launch of Mini-EUSO on the 22\textsuperscript{nd} August 2019.

\section{Conclusion}
\label{sect:conclusion}

We have developed a modular and extensible software for the Mini-EUSO instrument and future EUSO experiments based on the same data acquisition principle. The software has been tested successfully throughout the integration and qualification of the instrument. The Mini-EUSO instrument has now been launched to the ISS, but we will continue to build on this software for the upgrade of EUSO-TA, which is currently underway, and the upcoming EUSO-SPB2 NASA balloon flight in the next few years.

\acknowledgments 
We would like thank C. Fuglesang for useful discussions, E. Reali for engineering advice during the integration tests in Tor Vergata, M. Bertaina and his research group at the University of Turin for supporting the tests of Mini-EUSO at TurLab and C. Giammanco for contributions to the code. We also acknowledge the contributions of the JEM-EUSO collaboration to the Mini-EUSO project. The anonymous referees are thanked for their constructive reviews which helped to improve the paper. This work was partially supported by the Italian Ministry of Foreign Affairs and International Cooperation, Italian Space Agency (ASI) contract 2016-1-U.0, the State Space Corporation ROSCOSMOS via a contract between SINP MSU and RSC Energia, the Russian Foundation for Basic Research, grant \#16-29-13065, and the Alexandra och Bertil Gyllings Stiftelsen. S. Turriziani was an International Research Fellow of the Japan Society for the Promotion of Science. The documentation of the software was produced using Doxygen \cite{Doxygen}, Sphinx \cite{Sphinx}, Breathe \cite{Breathe} and Read the Docs \cite{RTD}. 

%%%%% References %%%%%

\bibliography{report}   % bibliography data in report.bib
\bibliographystyle{spiejour}   % makes bibtex use spiejour.bst

%%%%% Biographies of authors %%%%%

\vspace{2ex}\noindent\textbf{Francesca Capel} is a PhD student in Astroparticle physics at KTH Royal Institute of Technology in Stockholm, Sweden. Prior to beginning her PhD, she worked on a miniaturised radiation monitor at the European Space Agency as part of their Young Graduate Trainee program. Since 2015, she has been part of the Mini-EUSO team and the JEM-EUSO Collaboration. Her interests include the development of instrumentation for the detection of ultrahigh-energy cosmic rays, as well as searching for their sources using existing datasets from ground-based cosmic ray and neutrino detectors. 

\vspace{1ex}
\noindent Biographies of the other authors are not available.

%\listoffigures
%\listoftables
%\end{spacing}

\end{document}